\documentclass[acus]{pac99}

\usepackage{epsfig}
\newcommand{\bit}{\begin{Itemize}}
\newcommand{\eit}{\end{Itemize}}

\begin{document}
\title{Electro-optical Detection of Charged Particle Beams\thanks{Work 
supported in part by the U.S. Department of Energy under Contract
N0. DE-AC02-98CH10886.}} 

\author{Y.K.~Semertzidis\thanks{Email: semertzidis@bnl.gov},
V.~Castillo, R.C.~Larsen, D.M.~Lazarus, B.~Magurno\thanks{Deceased}, 
T.~Srinivasan-Rao,\\ 
T.~Tsang, V.~Usack, Brookhaven National Lab. \\
L.~Kowalski, Montclair State Univ. \\
D.E.~Kraus, Univ. of Pittsburgh}

\maketitle

\begin{abstract}
We have made the first observation of a charged particle beam by means 
of its electro-optical effect on the propagation of laser light in a
birefringent crystal at the Brookhaven National Laboratory
Accelerator Test Facility.  Polarized infrared light was coupled to a
LiNbO$_3$ crystal through a polarization maintaining fiber of 4
micron diameter. 

An electron beam 
in 10~ps bunches of 1~mm diameter was scanned across the
crystal.  The modulation of the laser light during passage of the
electron beam was observed using a photodiode with 45~GHz bandwidth.
The fastest rise time measured, 120~ps, was made in the single shot
mode and was limited by the bandwidth of the oscilloscope and the
associated electronics.  Both polarization dependent and polarization
independent effects were observed.  This technology holds promise of
greatly improved spatial and temporal resolution of charged particle
beams. 
\end{abstract}

\section{Introduction}

A collaborative effort has been initiated to develop an ultra-fast
charged particle detector based on the birefringence induced in an
optical fiber carrying polarized light due to the electric field of a
relativistic charged particle.  An analysis of such a detector is
described in~\cite{gls}.  The electro-optical effect in amorphous
optical media is known as the Kerr effect~\cite{kerr} and is quadratic
in the electric field $E$; $\phi = 2 \pi K E^2 d$, where $\phi$ is the 
ellipticity induced in the polarized light, $K$ is the Kerr
coefficient and $d$ is the length of the electric field region
experienced by the material.  In uniaxial crystals the induced
ellipticity is linear in the externally applied E-field and the effect 
is called Pockels effect~\cite{yariv}.  The induced phase delay is then given
by $\phi = 
\pi (V/V_\pi)$ with $V$ the applied voltage and $V_\pi$ the voltage
required for producing a phase shift equal to $\pi$ rad.  As a first
step towards realizing the single particle detector we used an
intense, short length, electron beam from the Accelerator Test
Facility (ATF) of Brookhaven National Lab (BNL) and the Pockels
effect to detect it by optical means. 

\section{Setup and sensitivity}

\begin{figure}[ht]
\centering
\epsfig{file=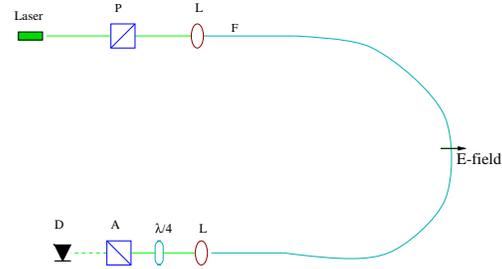, width=65mm}
\caption{The experimental setup for detecting charged particle beams
via optical means. 
The LiNbO$_3$ crystal was located at the beam position indicated by  E-field.
The positions of the polarizer (P), lenses (L), analyzer (A) and photodiode 
detector(D) are schematically indicated.}
\label{setup}
\end{figure}

The experimental setup in Fig.~\ref{setup} shows the laser (CW, from
Amoco  Laser Company) with 20~mW of optical power in the
infra-red ($\lambda = 1.32 \mu {\rm m}$), polarized by the polarizer
(P) and coupled to the fiber (F) with the microscope objective (L).
The fiber is polarization maintaining with a core of $4 \mu {\rm m}$
in diameter.

  The fiber was coupled to a commercially available LiNbO$_3$
crystal~\cite{utp} as indicated in Fig.~\ref{setup}. 
The crystal package was modified to allow for the 
passage of a charged beam without hitting the housing.  The laser
light was extracted from the fiber and its polarization state 
analyzed with by means of a $\lambda / 4$ plate and the analyzer (A).  
It was then detected by the photodiode and pre-amplifier~\cite{newfocus} 
the output of which goes to a fast transient digitizer~\cite{hp}.  

The transmitted light from the analyzer is equal to 

\begin{equation}
I = I_0 [ \sigma^2 + (\alpha + \phi(t))^2 ] \approx I_0 [ \sigma^2 +
\alpha^2 + 2 \alpha \phi(t) ] , \label{eq:intensity} 
\end{equation}

\noindent{with} $I_0$ the light intensity before the analyzer,
$\sigma^2$ the minimum possible ratio of ($I / I_0$) when $\alpha$ and 
$\phi$ are equal to zero, $\phi$ the induced ellipticity, and $\alpha$
is an intentional misalignment angle introduced to linearize and amplify 
the effect.  As is apparent from
Eq.~\ref{eq:intensity},
 the time dependent part of the light signal can 
be made positive or negative depending on the sign of $\alpha$
relative to $\phi$.

The ATF produced an electron beam of 45~MeV kinetic
energy, containing up to 1~nC in 10~ps bunches of 1~mm in diameter
and a repetition rate of 1.5~Hz.  This charged particle beam
creates an electric (E) field at a distance $r$ if $r\gg$ than the dimensions 
of the beam bunch:

\begin{equation}
E = \gamma N_e {q \over 4 \pi \epsilon_0 r^2 } = \gamma N_e \times 5.8\times 
10^{-5} {\rm V/m}, \label{eq:efield}
\end{equation}

{\noindent with} $\gamma$ the relativistic Lorentz factor, $N_e$ the number of
electrons in the beam, $q = 1.6 \times 10^{-19} {\rm C}$ the electron charge, 
$\epsilon_0 = 8.85 \times 10^{-12} {\rm F/m}$ the permittivity of free 
space, and $r$ the distance from the material (in the example we
used $r=0.5$~cm).  This electric field is present for

\begin{equation}
\Delta t = {r \over \gamma u} = {17 \over \gamma} {\rm ps} , \label{eq:time}
\end{equation}

{\noindent with} $u$ the beam velocity.

The  LiNbO$_3$ crystal used has $V_\pi = 5$~V with an electrode
separation of $15 \, \mu$m and a length of $l = 1.5$~cm.
The integral then of $\int E \, dl = \int {5V \over 15\times
10^{-6} \, {\rm m} } \, dl = 5000$~V is capable of producing $\pi$ rad 
of phase shift or $\pi / \sqrt{2}$ rad maximum of ellipticity (the
maximum ellipticity is induced when the laser polarization is at
$45^\circ$ with respect to the applied electric field direction).  The same
integral estimated for a particle beam 
located at the mid-plane orthogonal to the crystal at a distance $r=0.5$~cm 
is

\begin{equation}
\int E \, dl = { \gamma N_e q \over 4 \pi \epsilon_0 r } 2(1 -
\sin{\theta_1}) = 
\gamma  N_e \,  2.6\times 10^{-7} {\rm V} , \label{eq:ebeam}
\end{equation}

{\noindent with} $\theta_1$ the angle A\^{B}C where A is the
location of the beam, B one end of the crystal in the long direction
and C the center of the crystal.  This produces an ellipticity of 

\begin{equation}
\phi = \gamma N_e \times 0.1~{\rm nrad} , \label{eq:phase}
\end{equation}

{\noindent for} 17/$\gamma$~ps.  

The signal to noise ratio ($SNR$) for a detection system which is photon
statistics limited is given by 

\begin{equation}
SNR = \phi \sqrt{P T q_p \over 2 \hbar \omega}, \label{eq:snr}
\end{equation}

{\noindent with} $P$ the laser power, $T$ the inverse of the
detection system bandwidth, $q_p$ the quantum efficiency of the 
photodiode, and $\hbar \omega$ the energy of the laser photon.  As
an example we will take an electron particle beam with $\gamma = 1700$, 
$T=10$~fs, $q_p = 0.8$,  $P = 10^8$~W (e.g. 1~mJ pulsed laser
light for 10~ps) and $\hbar
\omega =  0.9$~eV, then the required number of electrons in the beam for 
$SNR = 1$ is $N_e \sim 2$.  

\section{Experimental results}

At the experimental setup we used a CW laser of 10~mW, and a detector
with 100~ps time resolution.  Then using the above
Eqs.~\ref{eq:time},~\ref{eq:snr} we estimated the number of particles
needed in 
the electron beam to be $N_e =  10^8$ for $SNR = 1$.  Our beam 
of 1~nC corresponds to $N_e = 9 \times 10^9$ which ensured its
detection.  The induced ellipticity (from Eq.~\ref{eq:phase}) is $\phi 
= 1.0\, \gamma$~rad equivalent to $\phi = 0.2$~rad when the signal
attenuation due to limited detector bandwidth is taken into account.

\begin{figure}[ht]
\centering
\epsfig{file=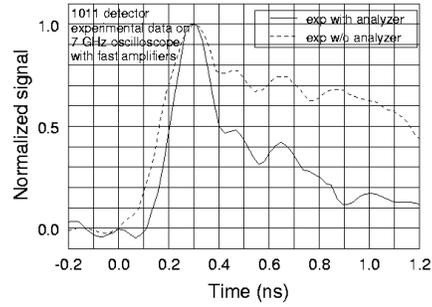, width=60mm,angle=-90}
\caption{The polarization dependent signal (solid line).  The electron beam 
was about $r=0.5$~cm from the crystal.  The polarization independent
signal is also shown (dash line). }
\label{signal1}
\end{figure}

In Fig.~\ref{signal1} we show the polarization dependent signal (solid 
line) as
observed with a single shot of the electron beam.  Changing the sign
of $\alpha$ (see Eq.~\ref{eq:intensity}) the signal also flips sign while 
retaining the same amplitude.  The maximum modulation of the light
intensity was about $9 \%$ of its DC level.

 We have also observed a signal when the crystal intercepted
the beam which is shown in Fig.~\ref{signal1} again in the single
shot mode (dash line).  We repeated this without the analyzer present and
found it was independent of it.  The difference from the polarization
dependent signal is two-fold.  First it does not flip sign under any
polarization orientation, and second it has a 
much longer time decay constant. 

We intend to improve the time resolution of the signal
by increasing the laser light intensity and the oscilloscope
bandwidth.  Long term plans include the implementation of this
sensitive method as a readout to a high rate single particle detector.

If $r$ is reduced to $100 \, {\rm \mu
m}$, the required $N_e$ is reduced by a factor of 2, making
possible ultra-fast single particle detection by optical means.
The gain factor is only 2 because of 
the limited bandwidth of the assumed detector which attenuates the
signal by the ratio of $ \Delta t / T$.

We will also look into a new type of fiber which exhibits high
polarizability~\cite{pole} thus reducing the cost of the detector considerably.


\begin{thebibliography}{9}

\bibitem{gls} Y.K. Semertzidis, XXVII International Conference on High 
Energy Physics gls0918 (1994).

\bibitem{kerr} J. Kerr, Phi. Mag. {\bf 50}, 337, 446 (1875).

\bibitem{yariv} A. Yariv, Quantum Electronics, Wiley, New York, 1967,
3rd ed. 1989.

\bibitem{utp} Uniphase Telecommunications Products, 1289 Blue Hills
Ave., Bloomfield, CT 06002.

\bibitem{newfocus} Type 1011 pre-amplifier from New Focus, Inc., 2630
Walsh Ave., Santa Clara, CA 95051.  The conversion gain of the
detector is 10~V/W, with a rise time of 9~ps. 

\bibitem{hp}  The oscilloscope used was from Hewlett Packard (HP SDC5000), 
of 7~GHz bandwidth.

\bibitem{pole} X.C. Long and S.R.J. Brueck, IEEE Photonic Technology
Letters, vol 9, p.767, 1997.

\end{thebibliography}
\end{document}